# Evolutionary Dynamics and Optimization

## Neutral Networks as Model-Landscapes for RNA Secondary-Structure Folding-Landscapes

Christian V. Forst*, Christian Reidys, and Jacqueline Weber

*Mailing Address:
Institut für Molekulare Biotechnologie
Beutenbergstraße 11, PF 100 813, D-07708 Jena, Germany
Phone: **49 (3641) 65 6459    Fax: **49 (3641) 65 6450
E-Mail: chris@imb-jena.de

## Abstract

We view the folding of RNA-sequences as a map that assigns a pattern of base pairings to each sequence, known as secondary structure. These preimages can be constructed as random graphs (i.e. the *neutral networks associated to the structure* $s$).

By interpreting the secondary structure as biological information we can formulate the so called *Error Threshold of Shapes* as an extension of Eigen's *et al.* concept of an error threshold in the single peak landscape [5]. Analogue to the approach of Derrida & Peliti [3] for a flat landscape we investigate the spatial distribution of the population on the neutral network.

On the one hand this model of a *single shape landscape* allows the derivation of analytical results, on the other hand the concept gives rise to study various scenarios by means of simulations, e.g. the interaction of two different networks [29]. It turns out that the *intersection* of two sets of compatible sequences (with respect to the pair of secondary structures) plays a key role in the search for "fitter" secondary structures.

# 1. Introduction

The first theory of biological evolution was presented last century by Charles Darwin (1859) in his famous book *The Origin of Species.* It is based on two fundamental principles, *natural selection* and *erroneous reproduction* i.e. *mutation.* The first principle leads to the concept *survival of the fittest* and the second one to *diversity*, where fitness is an inherited characteristic property of a species and can basically be identified with their *reproduction rate.*

Au contraire to Darwin's theory of evolution the role of stochastic processes has been stated. Wright [31, 32] saw an important role for genetic drift in evolution in improving the "evolutionary search capacity" of the whole population. He saw genetic drift merely as a process that *could* improve evolutionary search whereas Kimura proposed that the *majority of changes* in evolution at the molecular level were the results of random drift of genotypes [18, 19]. The neutral theory of Kimura does not assume that selection plays *no* role but denies that any appreciable fraction of molecular change is caused by selective forces. Over the last few decades however there has been a shift of emphasis in the study of evolution. Instead of focusing on the differences in the selective value of mutants and on population genetics, interest has moved to evolution though natural selection as an optimization algorithm on complex fitness landscapes. However, for a short moment let us return to Darwin and his minimal requirements for adaption:

- a population of object that are capable of replication,
- a sufficiently large number of variance of those objects,
- occasional variations which are inheritable, and
- restricted proliferation which is constrained by limited resources.

In this paper we restrict ourselves to RNA, the possibly simplest entities that do actually fulfill all the four requirements listed above. We realize the fundamental dichotomy of genotypic legislative by RNA and the phenotypic executive is manifested by RNA secondary structures. In this context the mapping from RNA sequences to secondary structures is of central importance, since fitness is evaluated on the level of structures. This mapping induces naturally a *landscape* on the RNA sequences independent of any possible evaluation of RNA structures [27]. Following the approach in [23] we can construct these sequence structure maps by random graphs. By omitting any empirical parameter of RNA-melting experiments we obtain the so called *neutral networks* of sequences which each fold into one single structure. It can be shown that these neutral networks and the transitions between them are "essential" structural elements in the RNA-folding landscape [24]. These landscapes combine both in the first view contradicting approaches on biological evolution; Darwins survival of the fittest and Kimuras neutral random drift.

# 2. Realistic Landscapes

## 2.1. Fitness Landscapes and the Molecular Quasispecies

In this contribution we consider the most simple example of Darwinian evolution,namely a population $\mathcal{P}$ of haploid individuals competing for a common resource.

Following the work of Eigen [4, 5] we consider a population of RNA sequences of fixed length $n$ in a stirred flow reactor whose total RNA population fluctuates around a constant capacity $N$. The definition of the overall replication rate of a sequence together with the constrained population size specifies our selection criterion.

In the limit of infinite populations its evolution is described by the *quasispecies equation*

$$\dot{c}_x = \sum_y Q_{xy} A_y c_y - c_x \Phi \tag{1}$$

where $c_x$ denotes the concentration of genotype $x$, $A_x$ is the replication rate of this genotype, and $Q$ is the matrix of mutation probabilities, $Q_{xy}$ being the probability for a parent of type $y$ to have an off-spring of type $x$. The replication rates considered as a function of the genotypes $x$ form a fitness landscape[1] [31] over the sequence space [6]. The total population is kept constant by a flux $\Phi$ compensating the production of new offsprings. The model mimics the asynchronous serial transfer technique [20].

As in the laboratory our RNA populations are tiny compared to the size of the sequence space. This fact forces a description in terms of stochastic chemical reaction kinetics. Two methods are appropriate to model stochastic processes:

- Gillespie [14] has described an algorithm for simulating the complete master equation of the chemical reaction network. We have used the implementation by Fontana *et al.* [10] for all computer simulations reported here. An individual sequence $I_k$ can undergo two stochastic reaction events: either $I_k$ is removed by the dilution flow, or it replicates with an average rate $A_k$ that corresponds to the reaction rate constant in equ. (1). When an individual sequence is replicated, each base is copied with fidelity $q$. The overall model mimics the asynchronous serial transfer technique [20].
- While giving an complete description of the dynamics Gillespies algorithm does not allow for a detailed mathematical analysis. Therefore we approximate the quasispecies model by a birth-death process, following the lines of Nowak & Schuster [21] and Derrida & Peliti [3]. All analytical results presented in this contribution are based on this approach.

---

[1] For a recent review on fitness landscapes see, e.g., the contribution by Schuster and Stadler in the proceedings volume to the Telluride meeting 1993 [25]

In general all rate and equilibrium constants of the replication process and hence also the over-all rate of RNA synthesis are functions of the 3D-structure.

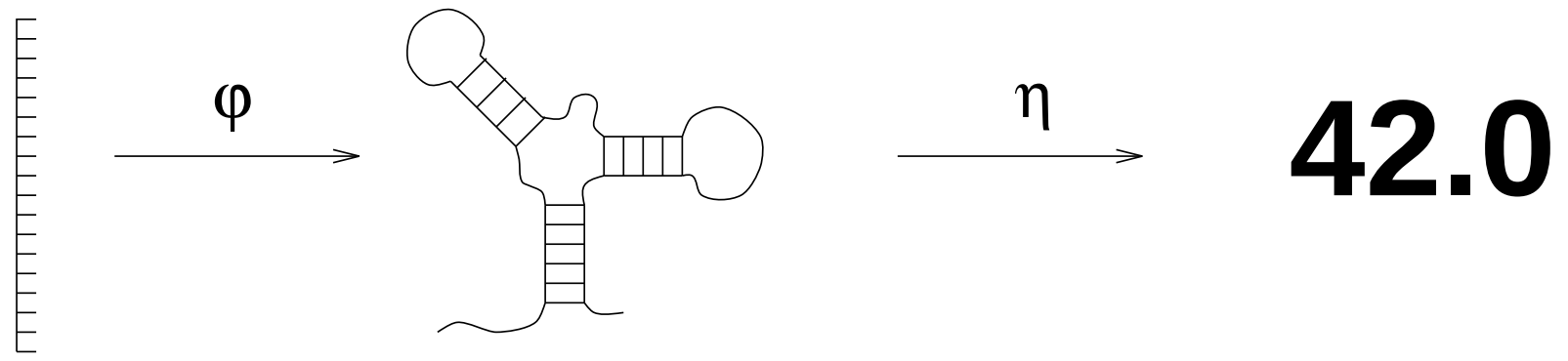

**Fig. 1.** Mapping of a genotype into its functional representation. The process is partitioned into two phases: The first phase is the complex mapping $\varphi$ of sequences into secondary structures (phenotypes). Here neutrality plays a crucial rôle; in the second phase we omit the building of the spatial 3D-structure and evaluating its function. We assign arbitrarily a fitness-value to each phenotype by the mapping $\eta$.

This suggests to decompose the computation of the fitness into two steps: First we construct the shape of the RNA (phenotype) from its sequence (genotype), and then we consider the evaluation of this phenotype by its environment (figure 1). The effect of this composition is that we are left with two hopefully simpler problems, namely (1) to model the relation between sequences and structures in the special case of RNA, and (2) to devise a sensible model for the evaluation of there structures. The *combinatory map* of RNA secondary structures, i.e., the map assigning a shape $\varphi(x)$ to each sequence in the sequence space $\mathcal{C}$ will be discussed in the next section.

Formally, we consider the evaluation $\eta$ assigning a numerical fitness value to each shape in the *shape space* $\mathcal{S}$. As even less is known in general about structure-function relations than about sequence-structure relations we will use the most simple model for the evaluation $\eta$. We assign arbitrary fitness-values $\eta(s_i)$ to specially chosen shapes $s_i$ and a fitness-value to the background $\eta(\beta)$ (i.e. the remaining shapes) with the condition $\eta(s_i) > \eta(\beta)$ for all $i$.

Tying things together we are considering a fitness landscape of the form

$$f(x) = \eta(\varphi(x)). \tag{2}$$

## 2.2. The Combinatory Map of RNA Secondary Structures

Having defined the evaluation $\eta$ of the structures we now turn to the sequence-structure relation $\varphi$. The phenotype of an RNA sequence is modeled by its minimum free energy (*MFE*) secondary structure.

The evidence compiled in a list of references [8, 9, 11, 15, 24, 26] shows that the combinatory map of RNA secondary structures has the following basic properties:

(1) Sequences folding into one and the same structure are distributed randomly in the set of "compatible sequences", which will be discussed below in detail.
(2) The frequency distribution of structures is sharply peaked (there are comparatively few common structures and many rare ones). Nevertheless, the number of different frequent structures increases exponentially with the chain length.
(3) Sequences folding into all common structures are found within (relatively) small neighborhoods of any random sequence.
(4) The shape space contains extended "neutral networks" joining sequences with identical structures. "Neutral paths" percolate the set of compatible sequences.
(5) There is a large fraction of neutrality, that is, a substantial fraction of all mutations leave the secondary structure completely unchanged (see figure 2).

These features are robust.

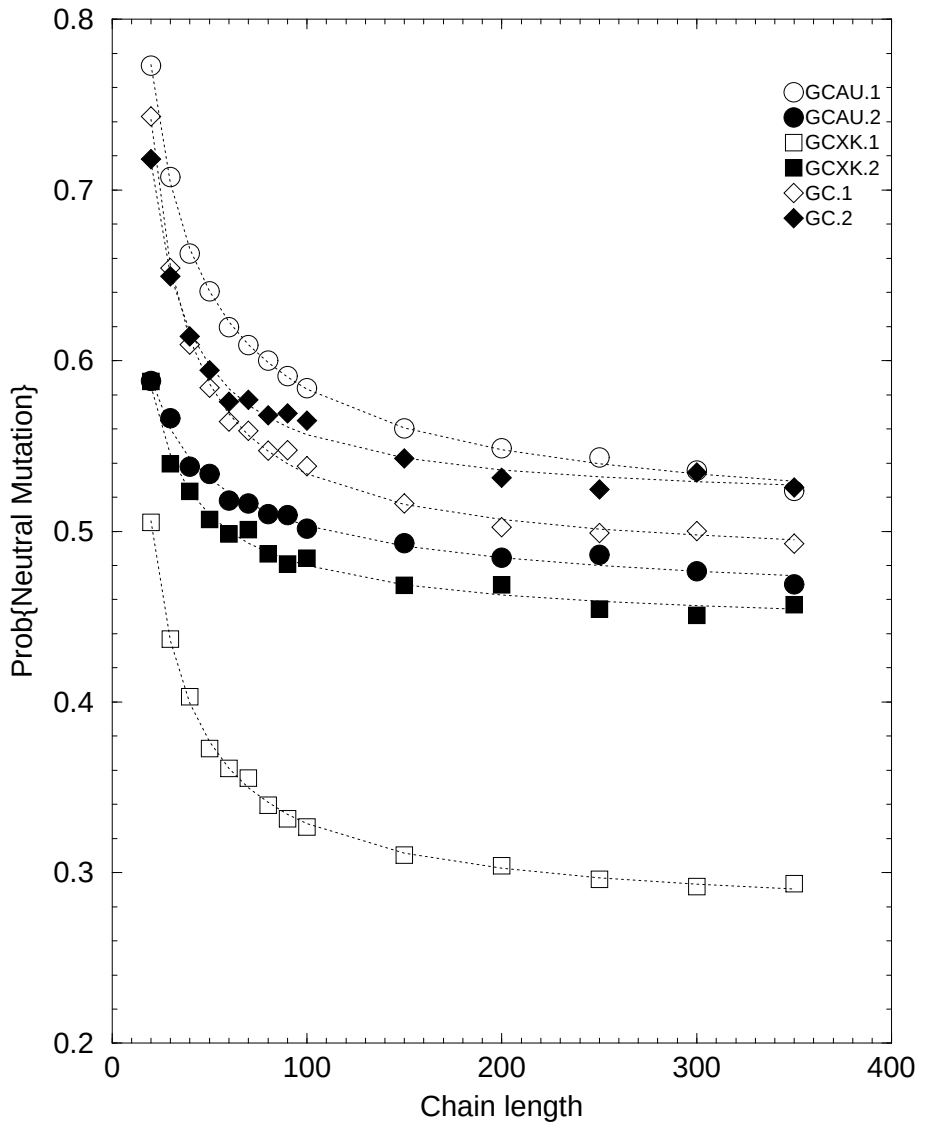

**Fig. 2.** Frequency of neutral mutations ($\lambda_u$ and $\lambda_p$ resp. — see section 2.3), counted separately for single base exchanges in unpaired regions (open symbols) and base pair exchanges (full symbols) for different alphabets.

A sequence $x$ is said to be *compatible* to a secondary structure $s$ if the nucleotides $x_i$ and $x_j$ at sequence positions $i$ and $j$ can pair whenever $(i, j)$ is a base pair in $s$. Note that this condition does by no means imply that $x_i$ and $x_j$ will actually form a base pair in the structure $\varphi(x)$ obtained by some folding algorithm. The set of all sequences compatible with a secondary structure $s$ will be denoted by $\mathbf{C}[s]$. There are two types of neighbors to sequence $x \in \mathbf{C}[s]$: each mutation in a position $k$ which is unpaired in the secondary structure $s$ leads again to a sequence compatible with $s$, while point mutations in the paired regions of $s$ will in general produce sequences that are not compatible with $s$. This problem can be overcome by modifying the notion of neighborhood. If we allow the exchange base pairs instead of single nucleotides in the paired regions of $s$ we always end up with sequences compatible with $s$. This definition of neighborhood allows us to view $x \in \mathbf{C}[s]$ as a graph. It can be shown [23] that this graph is the direct product of two generalized hypercubes

$$\mathbf{C}[s] = \mathcal{Q}_{\alpha}^{n_u} \times \mathcal{Q}_{\beta}^{n_p} \tag{3}$$

where $n_u$ is the number of unpaired positions in $s$, $\alpha$ is the number of different nucleotides, i.e., $\alpha = 4$ in the case of natural RNAs, $n_p$ is the number of base *pairs* in $s$, and $\beta$ is the number of different *types* of base pairs that can be formed by the $\alpha$ different nucleotides; for natural RNAs we have $\beta = 6$. The sequence length is $n = n_u + 2n_p$.

### 2.3. A Random Graph Construction

Folding RNA sequences into their secondary structures is computationally quite expensive. It is desirable, therefore, to construct a simple random model for the sequence structure map $\varphi$ with the same five properties that have been observed for RNA. Reidys *et al.* [23] have investigated random subgraphs of the hypercubes with the result that their approach is in fact able to explain the known facts about the combinatory map of RNA secondary structures.

#### *2.3.1. A Mathematical Concept*

We consider two closely related models. Consider a hypercube $\mathcal{Q}_{\alpha}^{n}$. We construct a random subgraph $\Gamma'_{\lambda}$ by selecting each edge of $\mathcal{Q}_{\alpha}^{n}$ independently with probability $\lambda$. From $\Gamma'_{\lambda}$ we obtain the induced subgraph $\Gamma^{\lambda} = \mathcal{Q}_{\alpha}^{n}[\Gamma'_{\lambda}]$ by adding all edges between neighboring sequences that have not been assigned already by the random process.[2] The probability $\lambda$ is simply the (average) fraction of neutral neighbors.

The main result about these random subgraph models is that there is a critical value $\lambda^*$ such that the subgraph $\Gamma_{\lambda}$ is dense in $\mathcal{Q}_{\alpha}^{n}$ and connected (i.e., for any two vertices $\Gamma_{\lambda}$ there is path in $\Gamma_{\lambda}$ that connects them) whenever $\lambda > \lambda^*$. Explicitly it has been shown [23] that

$$\lambda^* = 1 - \sqrt[1-\alpha]{\alpha}\,. \tag{4}$$

Density and connectivity of the neutral network $\Gamma$ result in percolating neutral paths.

[2] Alternatively, one could draw vertices from $\mathcal{Q}^n$ and consider corresponding the induced subgraph. Both random subgraph models have essentially the same properties.

*2.3.2. Modeling Generic Fitness Landscapes*

The model formulated above does not take into account that there are in general different probabilities for the two classes neutral mutations, $\lambda_u \neq \lambda_p$ for the *unpaired* and *paired* parts of the secondary structure, respectively. Using that the "graph of the compatible sequences" is a direct product of two hypercubes this limitation can be overcome by considering the direct product of two random graphs, one in each of the two hypercubes:

$$\begin{array}{ccccc} & & \mathcal{Q}_\alpha^{n_u} \times \mathcal{Q}_\beta^{n_p} & & \\ & & \uparrow \iota & & \\ & & \Gamma_{\lambda_u} \times \Gamma_{\lambda_p} & & \\ & \nearrow^{in} & & \nwarrow^{in} & \\ \Gamma_n^{\lambda_u} & & & & \Gamma_{\lambda_p} \end{array} \qquad (5)$$

This model inherits its properties from the basic random subgraph model on the two hypercubes. In particular $\Gamma = \Gamma_{\lambda_u} \times \Gamma_{\lambda_p}$ is dense and connected if both "components" $\Gamma_{\lambda_u}$ and $\Gamma_{\lambda_p}$ are dense and connected. From now on we will only refer to this model for deducing our results in this paper.

A neutral network induces in a natural way a *fitness landscape* $f_\Gamma$ on the complete sequence space $\mathcal{Q}_\alpha^n$:

$$f(x) = \left\{ \begin{array}{lll} 1 & \text{if} & x \notin \Gamma \\ \sigma & \text{if} & x \in \Gamma \end{array} \right\} . \qquad (6)$$

with $\sigma > 1$. We call $f_\Gamma$ a *single shape* landscape in contrast to the single peak landscapes discussed for instance in [4, 5]. The two degenerated cases $\lambda_u = \lambda_p = 0$ and $\lambda_u = \lambda_p = 1$ are referred to the single peak landscape ($\Gamma$ consists of a single sequence) and the *flat landscape* resp. In the following we will exploit the analogy between single peak and single shape landscapes quite extensively.

Summarizing the above discussion we claim that a single shape landscape is a much more realistic approximation of real fitness landscapes than a single-peak landscape or a spin glass like model landscape, since all these approaches lack what we think is the most important feature of biomolecular landscapes: *a high degree of neutrality.*

In chapter 5 we present a canonical generalization of the single-shape landscape to the more realistic *multi-shape* landscape. Transitions between two neutral networks are studied.

## 2.4. The Birth and Death Process Model

Let us now return to the dynamic behavior of a population $\mathcal{P}$ on such a landscape. Obviously $f_\Gamma$ induces a *bipartition* of the population $\mathcal{P}$ into the subpopulation $\mathcal{P}_\mu$ on the network $\Gamma$ and the remaining part $\mathcal{P}_\nu$ of inferior individuals. We call the elements of $\mathcal{P}_\mu$ *masters* (because they have superior fitness) and those of $\mathcal{P}_\nu$ *nonmasters.*

We will describe the evolution of $\mathcal{P}$ in $\mathcal{Q}_\alpha^n$ in terms of a *birth-death model* [17] with *constant* population size. At each step two individuals are chosen randomly; the first choice is subject to error-prone replication while the second choice is removed from the population [21]. The stochastic process is specified by the following probabilities:
$W^\Gamma_{\mu,\mu}$ is the probability that the offspring of a master is again a master;
$W^\Gamma_{\mu,\nu}$ is the probability that the offspring of a master is a non-master;
$W^\Gamma_{\nu,\mu}$ is the probability that the offspring of a non-master is a master; and
$W^\Gamma_{\nu,\nu}$ is the probability that the offspring of a non-master is again a non-master.

In general these probabilities will depend on the details of the surrounding of each particular sequence, namely on the number of neutral neighbors. It is possible, however, to show [23] that the fraction of neutral neighbors obeys a Gaussian distribution which approaches a $\delta$-distribution in the limes of long chains. The same behavior was found numerically for RNA secondary structures. Hence we can assume that the number of neutral neighbors is the same for all masters, namely $n\lambda_u$ and $n\lambda_p$ for the two classes of neighbors. Consequently the probabilities $W^\Gamma_{\mu,\mu}$, $W^\Gamma_{\mu,\nu}$, $W^\Gamma_{\nu,\nu}$, and $W^\Gamma_{\nu,\mu}$ are independent of the particular sequence.

We consider each replication-deletion event as one single event per time-step. The consequence of this assumption is that depending on the individual fitness the equidistant time-step $\Delta t$ in *reactor-time* results in different time-intervals per replication-round $\Delta\tilde{t}$ in physical time $\tilde{t}$ . I.e. master replicate $\sigma$-times faster than nonmaster yielding in $\sigma$-times more individuals per replicated master than per nonmaster per physical time step $\Delta\tilde{t}$. This difference between physical time $t$ and population-dependent *reactor time* $\tilde{t}$ has to be taken into account by calculating the probabilities for the replication-deletion events.

Analogously to the mutation-probabilities $W^\Gamma$ we setup four probabilities $P$:
$P_{\mu,\mu}$ is the probability for choosing a master for replication and deletion;
$P_{\mu,\nu}$ is the probability for choosing a master for replication and a nonmaster for deletion;
$P_{\nu,\mu}$ is the probability for choosing a nonmaster for replication and a master for deletion;
$P_{\nu,\nu}$ is the probability for choosing a nonmaster for replication and deletion.

For the so called *birth*- and *death*-probabilities we obtain $\mathbf{P}_{k,k+1} = P_{\mu,\nu}W^\Gamma_{\mu,\mu} + P_{\nu,\nu}W^\Gamma_{\nu,\mu}$ and $\mathbf{P}_{k,k-1} = P_{\mu,\mu}W^\Gamma_{\mu,\nu} + P_{\nu,\mu}W^\Gamma_{\nu,\nu}$ resp.

After some lengthy calculations [22] we are able to compute the stationary distribution $\boldsymbol{\mu}_p$ of the birth-death process. According to [7, 17] $\boldsymbol{\mu}_p$ is given by $\boldsymbol{\mu}_p(k) = \pi_p(k)/\sum_k \pi_p(k)$ Then the stationary distribution is completely determined by

$$\pi_p(k) = \frac{W^\Gamma_{\nu,\mu}}{\mathbf{P}_{k,k-1}} \frac{B(N,C_2)}{(k+C_1)\,B(1+C_1,k)\;B(N-(k-1),C_2)} \cdot \left[\frac{\sigma\,W^\Gamma_{\mu,\mu} - W^\Gamma_{\nu,\mu}}{W^\Gamma_{\nu,\nu} - \sigma\,W^\Gamma_{\mu,\nu}}\right]^{k-1} \tag{7}$$

where $B(x,y)$ is the Beta-function. $\Lambda_i$ and $C_i$ $(i=1,2)$ are defined as follows:

$$\Lambda_1 \stackrel{\text{def}}{=} \left[\sigma\,W^\Gamma_{\mu,\mu} - W^\Gamma_{\nu,\mu}\right] \quad \Lambda_2 \stackrel{\text{def}}{=} \left[W^\Gamma_{\nu,\nu} - \sigma\,W^\Gamma_{\mu,\nu}\right]$$

$$C_1 \stackrel{\text{def}}{=} \frac{(N-1)\,W^\Gamma_{\nu,\mu}}{\Lambda_1}, \text{ and } \quad C_2 \stackrel{\text{def}}{=} \frac{(N-1)\,\sigma\,W^\Gamma_{\mu,\nu}}{\Lambda_2}.$$

For the dynamics in physical times holds

$$\frac{W^\Gamma_{\nu,\mu}}{\mathbf{P}_{k,k-1}} = \frac{W^\Gamma_{\nu,\nu}((\sigma-1)(k-1)+N)}{k\,[\sigma(k-1)\,W^\Gamma_{\mu,\nu} + (N-k)W^\Gamma_{\nu,\nu}]}$$

## 3. Diffusion on “Neutral” Landscapes

In general, “diffusion” can be understood as movement of the *barycenter* of the population in the high-dimensional sequence-space via point-mutations. The barycenter $M(t)$ of a population at time $t$ is a real valued consensus vector specifying the fraction $x_{i\alpha}(t)$ of each nucleotide $\alpha \in \{$A,U,G,C$\}$ at every position $i$.

### 3.1. Diffusion in Sequence Space

Let us assume again that a secondary structure $s \in \mathcal{S}_n$ and its corresponding neutral network $\Gamma$ are fixed. In this section we study the *spatial distribution* of the strings on the network i.e. the spatial distribution of $\mathcal{P}_\mu$. Here we understand spatial distribution as distribution in *Hamming distances.* For this purpose we introduce the random variable $\hat{d}^\mu_\Gamma$ that monitors the pair distances in the population $\mathcal{P}$. The shape of the distribution of $\hat{d}^\mu_\Gamma$ is basically determined by the following factors.

- the distribution of the random variable $\hat{Z}_\mu$ whose states are the number of offspring.

- the *structure* of the neutral network $\Gamma$, given by the basic parameters for the construction of the random graph, $\{\lambda_u, \lambda_p, n_u, n_p\}$.
- the single digit error rate $p$ for the replication-deletion process.

We will assume in the sequel that $|\,\mathcal{P}_\mu\,| = \mathbf{E}[\hat{X}_p]$, in other words the number of strings located on the neutral network is *constant.* Taking into consideration the genealogies along the lines of Derrida & Peliti [3] we can express the probability of having different ancestors in *all* $i$ previous generations:

$$w_i \approx e^{-\mathbf{V}[\hat{Z}]\; i/(\mathbf{E}[\hat{X}_p]\,-1)}\,, \tag{8}$$

where $\hat{Z}$ describes the number of offspring produced by a master-string viewed as a random variable ($\mathbf{E}[\,.\,]$ and $\mathbf{V}[\,.\,]$ denote expectation value and variance resp.).

Following Reidys *et al.* [22] we consider then *random walks* on the neutral network $\Gamma$. For this purpose we introduce the probability $\varphi_\Gamma(t,h)$ of traveling a Hamming distance $h$ on $\Gamma$ by a random walk lasting $t$ generations.

In this section we restrict ourselves to alphabets consisting of complementary bases that admit only complementary base pairs (consider for example $\{\mathbf{G},\mathbf{C}\}$ or $\{\mathbf{G},\mathbf{C},\mathbf{X},\mathbf{K}\}$). We consider *moves* as point-events, i.e. each move occurs at precisely one time step $\Delta t$. By use of the regularity assumption, we obtain the *infinitesimal error rates* (for unpaired and paired digits), $\lambda_u\, p\, \Delta t$ and $\lambda_p\, p^2\, \Delta t$.

Arbitrarily we set

$$\wp_u(t) \overset{\text{def}}{=\!=} \frac{\alpha-1}{\alpha}\,(1-e^{-\frac{\alpha}{\alpha-1}\lambda_u\, p\, t}) \text{ and } \wp_p(t) \overset{\text{def}}{=\!=} \frac{\beta-1}{\beta}\,(1-e^{-\frac{\beta}{\beta-1}\lambda_p p^2\, t})\,. \tag{9}$$

Combining the information on the genealogies and the random walks allows us to compute the distribution of $\hat{d}^\mu_\Gamma$ and leads to the main result in this section. For an alphabet consisting of complementary bases with pair alphabet $\mathcal{B}$ we have

$$\begin{aligned}\boldsymbol{\mu}\{\,\hat{d}^\mu_\Gamma = h\,\} = \mathbf{V}[\hat{Z}] \sum_{h_u+2h_p=h} \int_0^\infty & B(n_u, \wp_u(2[\mathbf{E}[\hat{X}_p]\,-1]\,\tau), h_u) \\ & B(n_p, \wp_p(2[\mathbf{E}[\hat{X}_p]\,-1]\,\tau), h_p)\; e^{-\mathbf{V}[\hat{Z}]\,\tau}\, d\tau\,,\end{aligned} \tag{10}$$

where $\wp_u((2[\mathbf{E}[\hat{X}_p]\,-1]\,\tau), h_u), \wp_p((2[\mathbf{E}[\hat{X}_p]\,-1]\,\tau), h_p)$ are defined above.

Next we turn to the average distance between the populations $\mathcal{P}(t)$ and $\mathcal{P}(t')$, where $t \geq t'$ are arbitrary times. Then we mean by

$$\begin{aligned}\operatorname{dist}(\mathcal{P}(t),\mathcal{P}(t')) & \overset{\text{def}}{=\!=} \frac{1}{\mathbf{E}[\hat{X}_p]^{\,2}} \sum_{\substack{v\,\in\mathcal{P}_\mu(t)\\ v'\in\mathcal{P}_\mu(t')}} d(v,v') \\ \operatorname{avdist}(\mathcal{P}(t),\Delta t) & \overset{\text{def}}{=\!=} \langle \operatorname{dist}(\mathcal{P}(t'),\mathcal{P}(t'+\Delta t))\rangle_{t'}\,.\end{aligned} \tag{11}$$

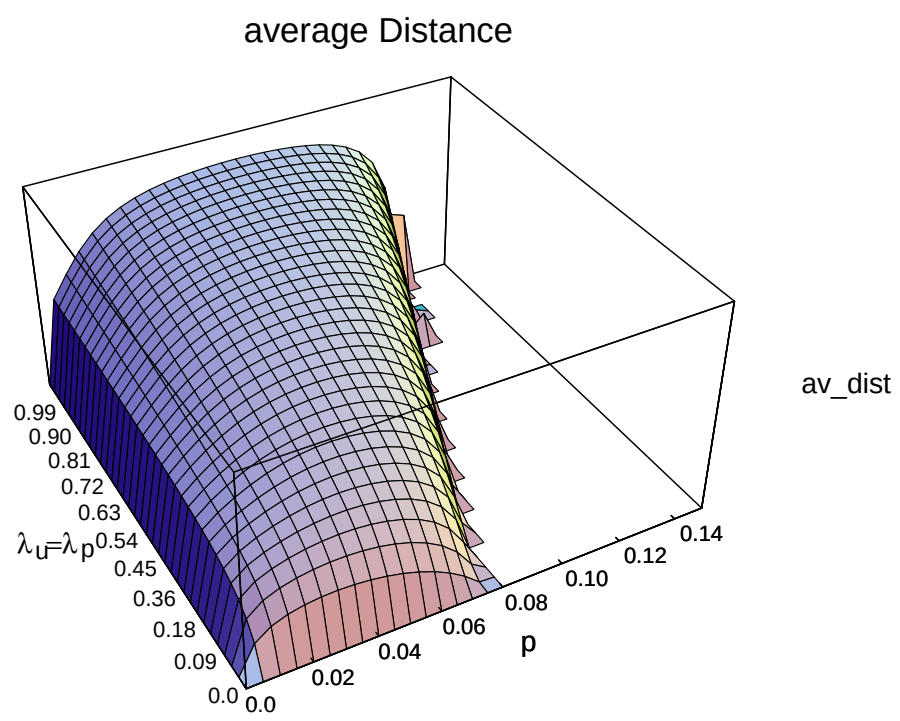

**Fig. 3.** The average pair distance $\mathbf{E}[\hat{d}^{\mu}_{\Gamma}]$ of master fraction of the population $\mathcal{P}$ on the neutral network $\Gamma$ in the long time limes. We assume that $\lambda = \lambda_u = \lambda_p$. The distance is plotted as function of the single digit error rate $p$ and the fraction of neutral neighbors for the paired and unpaired digits, $\lambda$. We observe that for wide parameter ranges the average pair distance of $\mathcal{P}_\mu$ is plateau-like. In particular the average pair distance becomes 0 at the shape-error threshold.

where $\langle\,.\,\rangle_{t'}$ denotes the *time average*. For binary alphabets with complementary base pairs it is shown in [22] that in the limes of infinite chain length

$$\begin{aligned} &\mathrm{avdist}[\mathcal{P}_\mu(t), \Delta t] \sim \\ &n_u/2 \left[\frac{\chi_u}{\chi_u + \mathbf{V}[\hat{Z}]}\right] [1 - e^{-2\lambda_u\, p\, \Delta t}] + n_p \left[\frac{\chi_p}{\chi_p + \mathbf{V}[\hat{Z}]}\right] [1 - e^{-2\lambda_p\, p^2\, \Delta t}] \end{aligned} \tag{12}$$

(see figure 3). Now we study the displacement of the *barycenter* of the population $\mathcal{P}_\mu$. For this purpose it is convenient to write the complementary digits $v_i$ of the sequence $x = (x_1, ..., x_n)$ as $-1$ and $1$ respectively. We write $x \cdot x' \stackrel{\mathrm{def}}{=} \sum_{i=1}^{n} x_i\, x_i'$.

The *barycenter* of the fraction of masters $\mathcal{P}_\mu \subset \mathcal{P}$ where $|\,\mathcal{P}_\mu\,| = \mathbf{E}[\hat{X}_p]$, denoted by $M^\mu(t)$, is

$$M^\mu(t) \stackrel{\mathrm{def}}{=} \frac{1}{\mathbf{E}[\hat{X}_p]} \sum_{v \in \mathcal{P}_\mu} v\,. \tag{13}$$

We can compute the resulting *diffusion-coefficient* $D$ of the barycenter $M^\mu(t)$ in the long time limes for a population $\mathcal{P}$ replicating on a neutral network $\Gamma$ with constant master fraction (implying a constant mean fitness $\overline{\sigma} = \frac{(\sigma-1)\mathbf{E}[\hat{X}_p]+N}{N}$). Explicitly the diffusion coefficient is given by

$$\begin{aligned} &\frac{1}{\Delta t}\, \langle [M(t+\Delta t) - M(t)]^2 \rangle_{t'} \approx \\ &2\,\lambda_u\, n_u\, p \left[\frac{\chi_u}{\chi_u + \mathbf{V}[\hat{Z}]}\right] + 4\,\lambda_p n_p p^2 \left[\frac{\chi_p}{\chi_p + \mathbf{V}[\hat{Z}]}\right]. \end{aligned} \tag{14}$$

and

$$\frac{1}{\Delta\tilde{t}}\langle[M(t+\Delta t)-M(t)]^2\rangle_{t'} = \overline{\sigma}\ \frac{1}{\Delta t}\langle[M(t+\Delta t)-M(t)]^2\rangle_{t'}\ ,$$

where $\chi_u = 4\,\lambda_u\,p\,(\mathbf{E}[\hat{X}_p]\,-1)$ and $\chi_p = 4\,\lambda_p\,p^2\,(\mathbf{E}[\hat{X}_p]\,-1)$.

### 3.2. Mutational Buffering

We can now compare the analytical distributions of $\hat{d}^{\mu}_{\Gamma}$ with our simulations done in the case of *binary* alphabets (see figure 4).

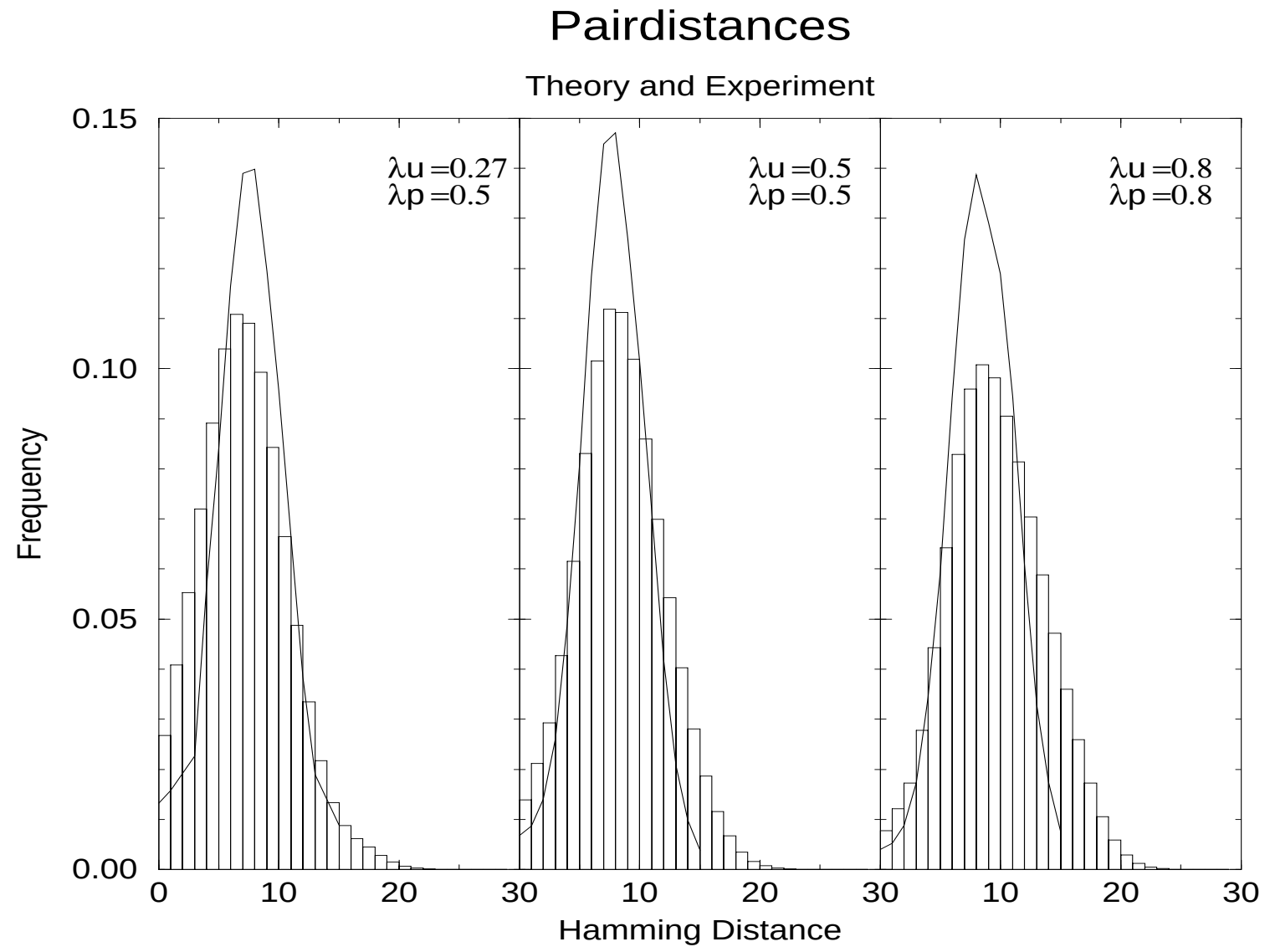

**Fig. 4.** The distribution of $\hat{d}^{\mu}_{\Gamma}$ in comparison to computer simulations that base on the Gillespie algorithm [14]. The simulation data are an time average for 300 generations. The solid lines denote the analytical values, the histograms show the numerical data.

The difference between the experimental and theoretical density curves is due to an effect known as *buffering* [16]. In the neutral networks a population is located preferably at vertices with higher degrees i.e.

$$v \in \mathrm{v}[\Gamma] : \ \delta_v \gg \lambda_u\,n_u + \lambda_p\,n_p\ .$$

For binary alphabets in particular the expected distance of pairs $(v, v')$ with $\delta_v, \delta_{v'} \gg \lambda_u\,n_u + \lambda_p\,n_p$ is $n/2$, since the distance sequence of the Boolean hypercube is given by $\binom{n}{k}$. Therefore we observe a shift to higher pair distances in the population as the theory predicts for regular neutral networks.

# 4. Phenotypic Error Threshold

## 4.1. Genotypic Error Threshold

We must distinguish between an error threshold with regard to the genotype (sequence) population and a different error threshold, at higher error rates, with regard to the induced phenotype (structure) population marking the beginning of drift in structure space. *That* is when the population can no more preserve the phenotypic information and optimization breaks down. In the present case this occurred[3] at $p \approx 0.1$ versus a sequence error threshold at approximately $p = 0$. What happens in between is, as it turns out, Kimuras neutral scenario [19] in a haploid asexually reproducing population.

## 4.2. Phenotypic Error Thresholds on the "Single Shape" Landscape

In this section we investigate the stationary distribution of the numbers of strings that are located on the neutral network $\Gamma$ (contained in a sequence space of fixed chain length $n$).

We shall discuss the following two extreme cases. On the one hand we can assume that the population size $N$ is infinite and on the other hand that $N \ll |\mathcal{Q}_\alpha^n|$. In the first case, since $n$ is assumed to be fixed, the concentrations of masters $c_\mu$ is *nonzero* for *all* error probabilities $p$.

Next let us consider the case $N \ll |\mathcal{Q}_\alpha^n| = \alpha^n$ i.e the population size is small compared to the number of all sequences. Since for any RNA secondary structures holds $n_p = O(n)$ and $n_u = O(n)$ we observe (for sufficiently large $n$) $\frac{|\Gamma|}{\alpha^n} \ll 1$. We now propose

$$p_N^* \stackrel{\text{def}}{=\!=} \max\left\{ p \mid \mathbf{V}[\hat{X}_p] = \left[\mathbf{E}[\hat{X}_p] - \frac{|\Gamma[s]|}{\alpha^n}\right]^2 \right\} \tag{15}$$

to be the *phenotypic error-threshold* for a population of $N$ strings replicating on a neutral network $\Gamma$. $p_\infty^*$ is further the *error threshold of the secondary structure* $s$. We immediately inspect that the above mentioned criterion generalizes the one used in the case of infinite population size in the single peak landscape of Eigen *et al.* [4], where $p_\infty^*$ is the solution of $c_\mu(p^*) = 1/\alpha^n$.

Let us discuss now the case of infinite population size. In this situation we can apply a completely deterministic ansatz solving a (well-known) rate equation

[3] depending on the fraction of neutral neighbors and relative superiority between individuals of different phenotype; a detailed study can be found i.e. in [12]

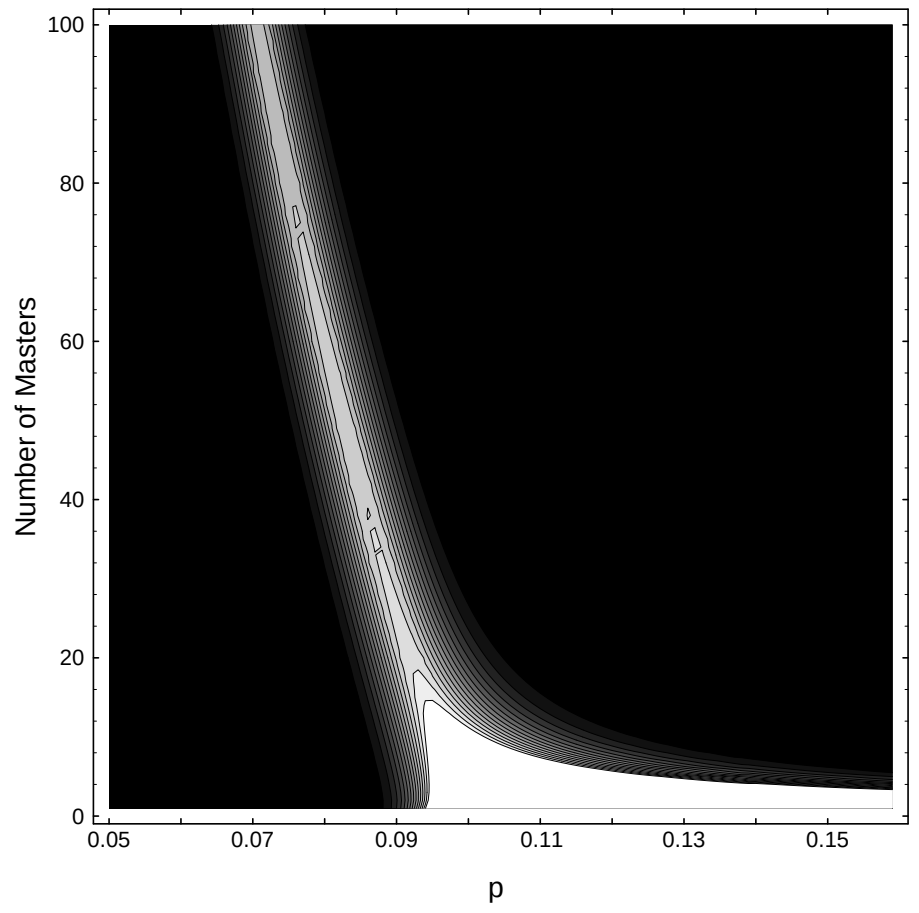

**Fig. 5.** For a regular neutral network $\Gamma$ with parameters $\lambda_u = 0.5$ and $\lambda_p = 0.5$ we plot the distribution of $\hat{X}_p$ i.e. the number of masters of $\mathcal{P}$.

for the corresponding *concentrations* of master $c_\mu$ and nonmaster vertices $c_\nu$, respectively. Assuming $W^{\Gamma}_{\nu,\mu} \approx 0$, i.e. neglecting back-flow mutations [4] and $\frac{|\Gamma|}{\alpha^n} \approx 0$, we derive

$$c_\mu \approx \left[\frac{\sigma W^{\tilde{\Gamma}}_{\mu,\mu} - (1 + W^{\tilde{\Gamma}}_{\nu,\mu})}{\sigma - 1}\right], \quad W^{\tilde{\Gamma}}_{\mu,\mu} \approx 1/\sigma \iff c_\mu = 0\,. \tag{16}$$

Using the threshold criterion of equ. (15) we can localize the error thresholds numerically for some population sizes and different Neutral-Network-landscapes.[4] with $\sigma = 10$ as superiority. The deterministic threshold values are obtained by solving $W^{\Gamma}_{\mu,\mu} \approx 1/\sigma$ (equ. (16)) for $p$ (table 1).

**Table 1:** Theoretical and numerical Error Thresholds (for $\sigma = 10$)

| $\lambda_u$ | $\lambda_p$ | Theory | | Experiment |
|---|---|---|---|---|
| | | $N = \infty$ | $N = 1000$ | $N = 1000$ |
| 0.1 | 0.1 | 0.079 | 0.071 | 0.065 |
| 0.27 | 0.5 | 0.081 | 0.08 | 0.0854 |
| 0.5 | 0.5 | 0.105 | 0.095 | 0.095 |
| 0.8 | 0.8 | 0.118 | 0.116 | 0.11 |

Finally we end this section by plotting the densities of the $i$-th incompatible classes $\mathbf{C}_i[s]$ (see figure 6) of the population obtained from our simulations[5]. We observe that at the error threshold there is a *sharp transition* from a population that is localized on the neutral network to a population that is uniformly distributed in sequence space.

[4]The calculations are done with *Mathematica* [30].

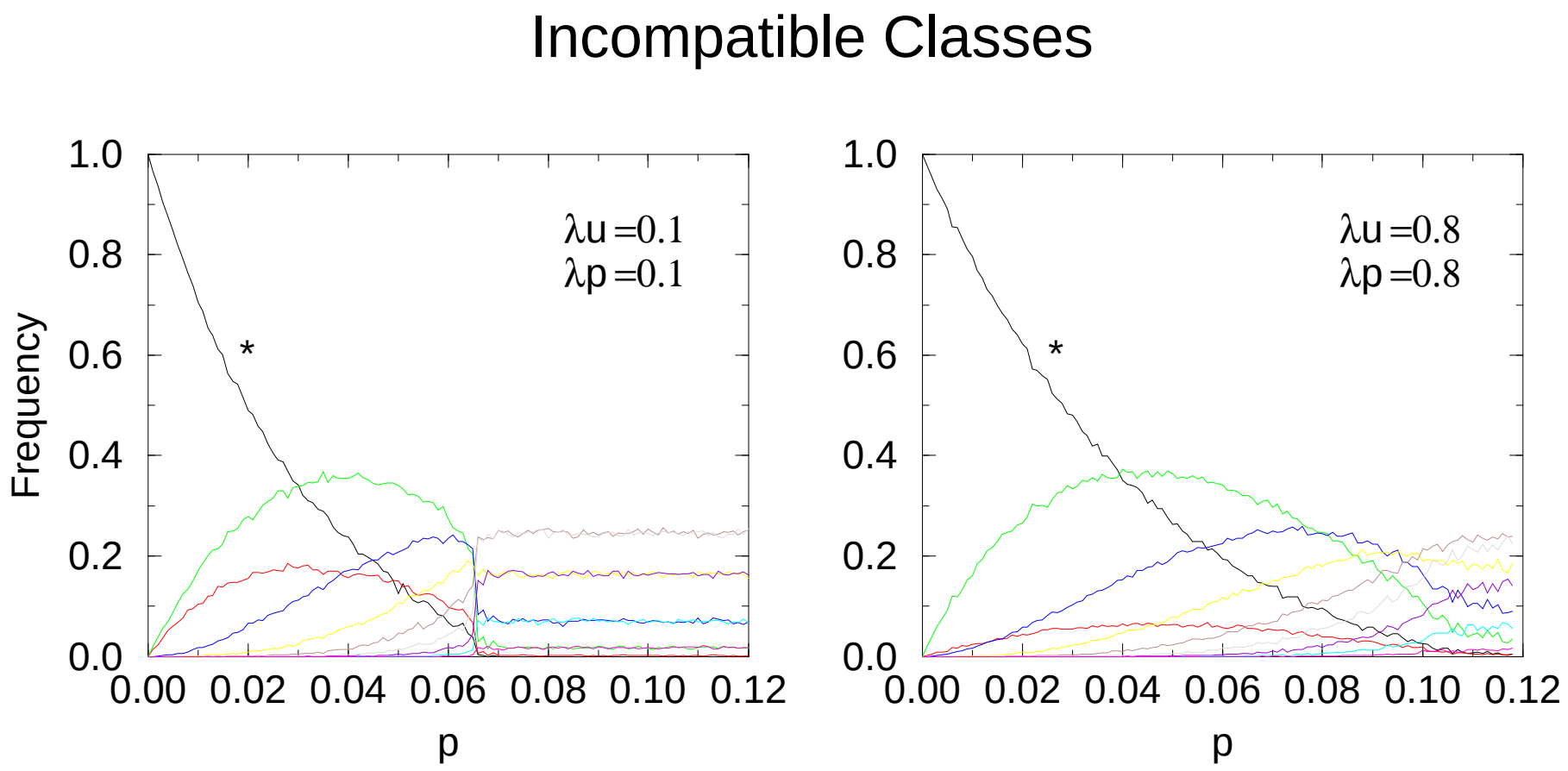

**Fig. 6.** In this figure we plot the error-classes in incompatible distances $\mathbf{C}_i[s]$ for certain Neutral-Network-landscapes. The underlying population size for the Gillespie simulation is $N = 1000$. The error-class $*$ denotes the number of masters i.e. the number of strings that are localized on the neutral network.

## 5. Transitions between Neutral Networks

Each neutral network is contained in the *set of compatible sequences* i.e. the set of sequences that could fold into one particular structure $s$. Each two sets of compatible sequences with respect to the pair of secondary structures have a nonempty *intersection*. This fact and the mathematical modeling of neutral networks as random graphs imply that the upper bound for the Hamming distance between two neutral networks is *four*. It turns out that the intersection is of particular relevance for *transitions* of finite populations of erroneously replicating strings between neutral neutral networks. In other words the intersection plays a key role in the search for "fitter" secondary structures.

It has been proven in [23] that the intersection is always nonempty. The intersection is constructed explicitly by using an algebraic representation of secondary structures. As already proposed in [23] each secondary structure $s$ can be interpreted as an element in $S_n$ by use of the mapping

$$\pi\colon\ \mathcal{S} \to S_n, \quad s \mapsto \pi(s) \stackrel{\text{def}}{=\!=} \prod_{i=1}^{n_p(s)} (x_i, x_i').$$

[5] In difference to the ansatz of constant population size, (the basic assumption for the birth-death model), the simulations are obtained by use of the Gillespie algorithm [14].

Here $(x_i, x_i')$ is a base pair in $s$ and $n_p(s)$ is the number of pairs in $s$. Clearly $\pi(s)$ is an *involution*, i.e. $\pi(s)\pi(s) = id$.

Using the fact that any two involutions $\imath$, $\imath'$ form a *dihedral group* $D_m \stackrel{\text{def}}{=\!=} \langle \imath, \imath' \rangle$ for secondary structures $s$ and $s'$ this leads to the mapping

$$j\colon \mathcal{S} \times \mathcal{S} \to \{D_m < S_n\}, \quad j(s, s') \stackrel{\text{def}}{=\!=} \langle \pi(s), \pi(s') \rangle.$$

In fact the operation of $\langle \pi(s), \pi(s') \rangle$ and especially the corresponding cycle decomposition is closely related to the structure of the intersection $\mathbf{I}[s, s']$.

An arbitrary cycle $z$ is given by a sequence of positions where predecessor and successor are determined by the pairs in $s$ and $s'$. We distinguish between two types of cycles: *open* and *closed* ones.

### 5.1. Size of the Intersection

This knowledge enables us to determine the size of the intersection. For alphabets $\mathcal{K}$ with complementary base pairs, e.g. $\mathcal{A} = \{G, C, X, K\}$ with corresponding base pairings $\mathcal{B} = \{GC, CG, XK, KX\}$, and $\alpha = |\mathcal{A}|$ we obtain

$$|\,\mathbf{I}[s, s']\,| = \alpha^{n_1 + n_2 + \ldots + n_r},$$

where $n_i$ is the number of cycles of length $i$. If we consider the physical alphabet $\mathcal{A} = \{G, C, A, U\}$ with corresponding pair alphabet $\mathcal{B} = \{GC, CG, AU, UA, GU, UG\}$ we obtain

$$|\,\mathbf{I}[s, s']\,| = \prod_{i=1} (\alpha_i^{(o)})^{n_i^o}\, (\alpha_{2i}^{(c)})^{n_{2i}^c}$$

where $n_i^c$ is the number of closed cycles with length $i$ and $n_i^o$ is the number of open cycles of length $i$ and $\alpha_\nu^{(o)}$ and $\alpha_\nu^{(c)}$ are the numbers of all possible configurations for an open cycle of length $\nu$ or a closed cycle of length $\nu$ with

$$\alpha_\nu^{(o)} = \frac{2}{\sqrt{5}} \left[ \left( \frac{2}{-1+\sqrt{5}} \right)^{\nu+2} - \left( \frac{2}{-1-\sqrt{5}} \right)^{\nu+2} \right] \quad \text{and}$$

$$\alpha_\nu^{(c)} = \frac{4}{\sqrt{5}} \left[ \left( \frac{2}{-1+\sqrt{5}} \right)^{\nu-1} - \left( \frac{2}{-1-\sqrt{5}} \right)^{\nu-1} \right]$$

.

### 5.2. Structure of the Intersection

**Definition 1.** *Let $s$ and $s'$ be two secondary structures. The graph $\mathcal{I}[s,s']$ has the vertex set $\mathbf{I}[s,s']$. Two sequences $x,y \in \mathbf{I}[s,s']$ are neighbors, e.g. $\{x,y\} \in \mathbf{e}[\mathbf{I}[s,s']]$, if and only if $x,y$ are neighbors in $\mathcal{C}[s]$ and $\mathcal{C}[s']$*

That means the intersection graph can be directly embedded in the graph structure of the sets of compatible sequences [23]. Note that the common unpaired positions in $s$ and $s'$ are the elements in the cycles of length 1. The common pairs of $s$ and $s'$ are represented by the closed cycles of length 2. Thus there exist two scenarios

(1) There are only open cycles of length 1 and closed cycles of length 2, then $\mathcal{I}[s,s']$ is a connected graph.

(2) There is at least one cycle of length greater than or equal 3 or one open cycle of length 2, then $\mathcal{I}[s,s']$ decomposes into components of equal size $((\alpha_1^{(o)})^{n_1^o} \cdot (\alpha_2^{(c)})^{n_2^c})$. The components are connected by paths in $\mathcal{C}[s]$ and $\mathcal{C}[s']$.

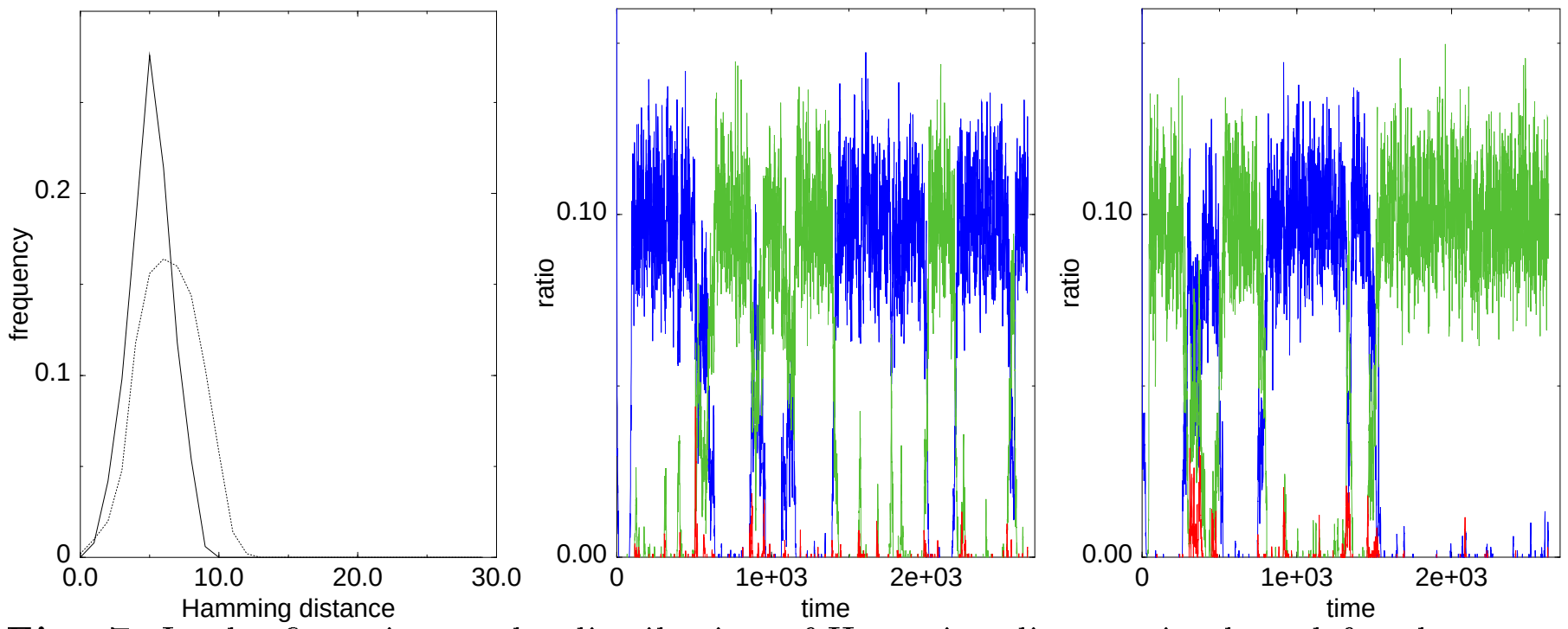

**Fig. 7.** In the first picture the distribution of Hamming distance is plotted for elements on $\mathcal{C}[s]$ to the intersection $\mathbf{I}[s,s_1']$ (solid line) and to $\mathbf{I}[s,s_2']$ (dotted line) resp. The second and the third picture show the Gillespie simulations, assuming that there is a same high fitness for both neutral networks and a low fitness elsewhere. Obviously the population uses the intersection to move from one network to the other.

### 5.3. Numerical Results

Suppose there are given two pairs of structures $(s,s_1')$ and $(s,s_2')$. We assume all $\lambda$ values to be equal and an action probability of $1/2$ on the intersection (figure 7).

The numerical results confirm the basic assumption of the neutral theory of Motoo Kimura [19]. The fixation of phenotypes is a consequence of a stochastic process.

## 6. Conclusions

Doing evolutive optimization on RNA secondary-structure folding landscapes is somehow different to optimization on typical rugged fitness landscapes. There are no local optima in the naive sense, but rather extended labyrinths of connected equivalent sequences which somewhere touch or come close to labyrinths of better sequences [13, 29]. What looks like punctuated equilibria in one projection (phenotype), presents itself as relentless and extensive change in another projection such as genetic makeup. Seen from this perspective the replicator concept that views genes as the sole unit of selection [2] may need an overhaul, since phenotypes are here to stay much longer than genes [1, 28].

Additional constraints at the sequence and the structure level may severely restrict the extent of neutral networks. However, there is no doubt about the evolutionary implications, should it turn out that RNA structures capable of performing biochemically interesting tasks do form neutral networks in sequence space or can be accessed from such networks. Given present day *in vitro* evolution techniques, these issues are within reach of experimental investigation.